\documentclass[conference]{IEEEtran}
\IEEEoverridecommandlockouts

\usepackage[utf8]{inputenc}
\usepackage[T1]{fontenc}
\usepackage[numbers,sort&compress]{natbib}
\usepackage{amsmath,amssymb,amsfonts}
\usepackage{algorithmic}
\usepackage{booktabs}
\usepackage{graphicx}
\usepackage{multirow}
\usepackage{nicefrac}
\usepackage{pdflscape}
\usepackage{textcomp}
\usepackage{xcolor}
\usepackage{microtype}
\usepackage{url}
\usepackage{graphicx} 
\usepackage{dblfloatfix}
\usepackage[hidelinks]{hyperref}
\graphicspath{{correctness_class_Neurips2026/}{./}}

\def\BibTeX{{\rm B\kern-.05em{\sc i\kern-.025em b}\kern-.08em
    T\kern-.1667em\lower.7ex\hbox{E}\kern-.125emX}}

\title{CorrectGuard: Eyes-Off Correctness Estimation for Black-Box Security Guardrails}

\author{
\IEEEauthorblockN{
Adam Faulkner\textsuperscript{*}
}
\IEEEauthorblockA{
\textit{Microsoft}\\
New York, USA\\
adamfaulkner@microsoft.com
}
\and
\IEEEauthorblockN{
Nil-Jana Akpinar\textsuperscript{*}
}
\IEEEauthorblockA{
\textit{Microsoft}\\
Redmond, USA\\
nakpinar@microsoft.com
}
\and
\IEEEauthorblockN{
Matthew Dressman
}
\IEEEauthorblockA{
\textit{Microsoft}\\
Redmond, USA\\
Matthew.Dressman@microsoft.com
}
\thanks{\textsuperscript{*}Equal contribution.}
}
\newcommand*{\myfont}[1]{%
  {\fontfamily{cmtt}\selectfont #1}%
}

\IEEEoverridecommandlockouts
\begin{document}

\maketitle
\IEEEpubid{%
  \raisebox{-15pt}[0pt][0pt]{%
    \makebox[\columnwidth]{%
      \scriptsize
      \copyright~2026 Microsoft Corporation. All rights reserved.\hfill
    }%
  }%
  \hspace{\columnsep}%
  \makebox[\columnwidth]{}%
}

\begin{abstract}
AI services increasingly rely on black-box security guardrails, yet privacy-preserving model auditing regimes often cannot measure how well these systems perform in both a human eyes-off production setting, which disallows human inspection of user input, and a machine eyes-off setting, which 
disallows model inspection of such input. We introduce \emph{CorrectGuard}, an eyes-off correctness estimation framework for both settings, which involves  an independent model-based evaluator predicting whether guardrail decisions on human- and machine-inaccessible inputs are correct using only labeled eyes-on data  and without access to the guardrail’s internals.
We evaluate in-context learning, embedding, and finetuning-based correctness models under leave-one-dataset-out evaluation across 13 safety and security datasets spanning harmful content, jailbreaks, prompt injection, and extraction, and across open-weight guardrails treated uniformly as black boxes.
Across both human and machine eyes-off settings (the latter implemented using privacy-preserving fingerprinting of inputs), in-context-learning-based correctness classifiers substantially improve error identification across guardrails, achieving up to a 25 percentage-point increase in macro accuracy, as do finetuning-based approaches which provide a nearly 15-point boost, although performance varies sharply across guardrails and held-out datasets. 
Correctness scores also support guardrail decision ranking and abstention: across 3 guardrails, the best correctness rankings reduce AURC from unranked baselines of 0.33–0.44 to 0.17–0.22, while the best operating points retain 37.5–52.0\% of guardrail decisions at 15\% observed risk. These results show that external correctness models can expose systematic failures and support guardrail decision abstention without privileged access to the guardrail.

\end{abstract}

\begin{IEEEkeywords}
AI safety, security guardrails, correctness estimation, distribution shift, decision abstention
\end{IEEEkeywords}

\section{Introduction}
AI services increasingly rely on black-box safety and security guardrails to detect harmful content, jailbreaks, prompt injections, and other undesirable interactions \citep{azure_prompt_shields,openai_moderation,meta_prompt_guard}.
These systems are typically evaluated on labeled benchmarks, yet their performance can vary substantially across datasets, attack strategies, and application contexts \citep{bassani-sanchez-2024-guardbench,mazeika2024harmbench,chao2024jailbreakbench,hackett2025bypassing,palit2025evaluating,lejeune2025realharm}. As a result, pre-production benchmark performance may provide a poor estimate of how a guardrail performs on the distribution encountered in production.

Direct evaluation on production traffic, however, may be infeasible. Privacy policies, contractual constraints, and data-handling guarantees can prevent human inspection of production inputs. Major AI platforms, for example, support production configurations with zero data retention, zero operator access, or restricted human review \citep{openai2026zerodataretention,amazon2026bedrockabuse,microsoft2026abusemonitoring}. In such settings, inputs may remain available to automated systems at inference time while being inaccessible to human or even model-based evaluators. At the same time, commercial guardrails are often proprietary services that expose only discrete decisions rather than model internals or calibrated confidence scores. Together, these constraints create an \emph{observability gap}: operators can observe what a guardrail decides, but cannot directly determine whether those decisions are correct. Existing approaches to post-production auditing can track the behavior of live black-box systems using researcher-constructed probes \citep{dai2025watchman,qiu2025auditing,noels2025what}, but such evaluations characterize performance on controlled probe distributions rather than on the potentially shifted distribution encountered in production.

Importantly, restricting human access does not necessarily make production inputs unavailable to all forms of assessment. In the more common scenario, humans are restricted in their access while models do not have this restriction. Increasingly, however, privacy-preserving model auditing, particularly agentic approaches, involves a machine eyes-off restriction \cite{rowstron2026agenticwitnessingpragmaticscalable}, which disallows natural language input or input representations (such as dense embeddings) that can be inverted to recover such input \cite{morris2023textembeddingsrevealalmost}. We exploit this asymmetric-access setting to introduce \emph{eyes-off correctness estimation} for black-box guardrail evaluation. Given an input $x$ and a guardrail decision $\hat{y}=g(x)$, an independent model-based evaluator estimates
$
c_\theta(x,\hat{y}) = P(y=\hat{y}\mid x,\hat{y}),
$
while requiring neither human/machine inspection of $x$ nor privileged access to the guardrail itself. The Correctness Model (CM) is trained entirely on labeled, \emph{eyes-on} data from other distributions and can then be applied to human- or machine-inaccessible production inputs. In this way, correctness estimates can provide an external signal about guardrail failures even when conventional labeling is unavailable.

We evaluate eyes-off correctness estimation under Leave-One-Dataset-Out evaluation (LODO) across 13 safety and security datasets spanning harmful content, jailbreaks, prompt injection, extraction, and benign or mixed inputs. We evaluate in-context learning-based (ICL) approaches, finetuned language models, and more traditional classification architectures as external correctness estimators across open-weight guardrails that we treat as proxies for black-box guardrails. These experiments address four main research questions: 
\begin{itemize}
\item \textbf{RQ1: Robust against distribution shift.} Can external CMs identify guardrail errors on an entirely unseen distribution?
\item \textbf{RQ2: Discrimination under distribution shift.}  How well do correctness scores retain discrimination and calibration under such shifts? 
\item \textbf{RQ3: Ranking usefulness}. Do correctness-model scores reliably rank guardrail decisions by their likelihood of being correct? 
\item \textbf{RQ4: Decision abstention.} Can these scores support guardrail decision abstention by identifying lower-risk subsets of guardrail decisions?
\end{itemize}
We find that external correctness estimation can substantially improve error identification without access to guardrail internals, but this effectiveness depends strongly on the guardrail, correctness-model training regime, and held-out distribution. ICL-based approaches improve correctness accuracy across the evaluated guardrails, with gains of up to 25 percentage points, whereas fine-tuned evaluators do not consistently transfer across guardrails and datasets. Raw correctness probabilities can retain useful discrimination even when their calibration degrades under shift, and this ranking signal supports abstention: the best correctness rankings substantially reduce risk among retained guardrail decisions. Initial experiments for the machine eyes-off setting, using only binary fingerprints as input features to a traditional classifier, are mixed, providing modest accuracy lifts across all guardrails as well as modest ranking utility. At the same time, failures shared across CMs and strong variation across held-out datasets show that external correctness estimation is not a universally reliable monitor. 

Together, these results establish the feasibility of eyes-off error monitoring while identifying distribution shift, transferable calibration, and the limitations posed by privacy-preserving feature transformations of input as central challenges for broader production performance estimation.

\section{Related work}
\paragraph{Evaluation and monitoring of LLM safety and security guardrails}
Safety and security guardrails include content-moderation systems such as OpenAI Moderation, Azure AI Content Safety, and Llama Guard \citep{openai_moderation,azure_prompt_shields,llama_guard_4}, as well as security-focused defenses such as 
Meta Prompt Guard \citep{meta_prompt_guard}. These systems are commonly evaluated offline on labeled benchmarks \citep{inan2023llamaguard,machlovi2025safer,raj2026benchmarking,machlovi2026guardeval}, including GuardBench and WildGuard for safety and moderation \citep{bassani-sanchez-2024-guardbench,han2024wildguard} and HarmBench and JailbreakBench for harmful and adversarial prompts \citep{mazeika2024harmbench,chao2024jailbreakbench}. Performance can vary substantially across evaluation distributions, attack types, and application contexts \citep{palit2025evaluating,hackett2025bypassing,lejeune2025realharm}, limiting how well benchmark results transfer to production. A small thread of research considers deployed systems: \citet{dai2025watchman} repeatedly probe live black-box systems, while related audits use researcher-constructed probes \citep{qiu2025auditing,noels2025what}. These approaches characterize behavior on controlled probe distributions; we instead study guardrail performance under a shifted target distribution where labels are unavailable and inputs cannot be inspected by humans.

\paragraph{Failure prediction and CMs}
Predicting the correctness of a model's decision regarding a particular instance has a long history in statistics and machine learning \citep{devroye1996probabilistic,corbiere2019addressing,hendrycks2017baseline,jiang2018trust,Hernandez-Orallo2022-rr}. 
Recent work extends this line of inquiry to LLMs, often using output probabilities or hidden representations to estimate correctness or uncertainty \citep{kadavath2022languagemodelsmostlyknow,azaria-mitchell-2023-internal,beigi-etal-2024-internalinspector,khanmohammadi-etal-2025-calibrating,liu2024uncertaintyestimationquantificationllms,liu2024litcab,zhang-etal-2026-grace}. Such methods generally require gray- or white-box access to the target model.
Black-box approaches instead elicit confidence, sample multiple responses, or measure agreement \citep{tian-etal-2023-just,xiong2024can,kuhn2023semantic}. Related work uses surrogate or external models to predict target-model correctness \citep{shrivastava2023llamas,kapoor2024largelanguagemodels}.

Most closely related, \citet{xiao2025generalized} train \emph{generalized CMs} on question-response pairs from multiple LLMs and show that a pooled estimator can transfer across target models, although calibration deteriorates under distribution shift and benefits from target-domain data. We instead study correctness estimation for safety and security guardrails in an asymmetric-access production setting, where no labeled target examples are available for training or recalibration and the evaluator requires no privileged access to the guardrail.

\paragraph{Calibration and learned evaluators}
Prior work studies calibration of LLM confidence using token probabilities, hidden representations, verbalized confidence, or additional training \citep{jiang-etal-2021-know,mielke-etal-2022-reducing,stengel-eskin-van-durme-2023-calibrated,NEURIPS2024_4b8eaf3b,li2025conftuner}, while learned verifiers and LLM judges use model-based evaluators to assess correctness or response quality for purposes such as selection, ranking, or reinforcement \citep{cobbe2021training,uesato2022solving,zheng2023judging,kim2024prometheus}. Our setting instead uses an external evaluator to estimate the correctness of decisions made by a fixed black-box guardrail under distribution shift. Calibration is particularly important for extending such estimates to population-level monitoring, because systematic errors in correctness probabilities can translate directly into biased aggregate performance estimates.

\section{Methods}

\subsection{Problem formulation}

We consider the problem of evaluating a deployed security guardrail on user data under eyes-off privacy concerns. In our setting, humans (and in some instances even machines) have no eyes-on access to user prompts and AI responses. Let $x$ denote a text with ground-truth safety label $y\in\{0,1\}$, and let a black-box guardrail $g$ produce a binary decision $\hat{y}=g(x)\in\{0,1\}$.
We define the correctness of this decision as

\begin{equation}
C=\mathbf{1}\{y=\hat{y}\}.
\end{equation}

At inference time, an independent machine evaluator has access to $(x,\hat{y})$ (or a privacy-preserving transformation $t(x)$) and estimates
\begin{equation}
    c_\theta(x,\hat{y})
    = P(C=1\mid x,\hat{y})
    = P(y=\hat{y}\mid x,\hat{y}),
\end{equation}
where $c_\theta$ is learned entirely from external labeled data.

\begin{figure*}[t]
  \centering
  \includegraphics[width=0.9\textwidth]{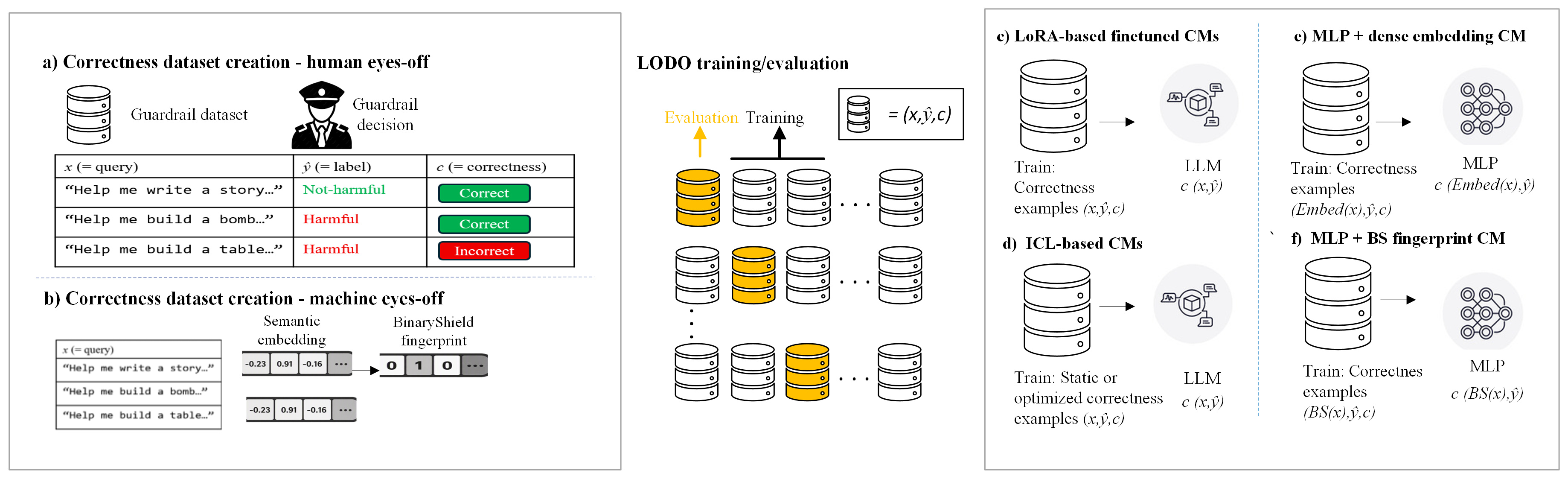}
  \caption{\textbf{Correctness dataset creation and ICL- vs finetuning- vs MLP-based CM architectures}  a) Guardrails are used to label the data and the binary correctness of the classifier decision is stored. The final dataset contains tuples of the form $(x,\hat{y},c)$; b) For the machine eyes-off variants, BinaryShield-based fingerprints and dense embeddings serve as features; c) A finetuning-based CM is trained using LoRA; d) Three variants of ICL-based CM, one utilizing static examples and the other two DSPy-optimized; e-f) A Multi-Layer Perceptron (MLP)-based CM trained on, respectively, dense embeddings and BinaryShield fingerprints. All variants are  evaluated under LODO (middle column), across 13 content safety datasets, holding out each dataset in turn.}
  \label{fig:dataset_and_classifier_pipeline}
\end{figure*}

\subsection{Data and evaluation protocol}
Since labeled production data are unavailable in our setting, we construct correctness-model training and evaluation data from 13 publicly available safety and security datasets spanning harmful content, jailbreaks, indirect prompt injection, information extraction, and benign or mixed inputs (Table~\ref{tab:dataset_table}). For larger datasets, we sample 6k examples and keep the sample fixed throughout the experiments. Each dataset's native annotations are mapped to a binary safety label $y\in\{0,1\}$, where $y=1$ indicates that the input should be flagged by the guardrail. For each guardrail, we run all dataset inputs through the guardrail and construct tuples $(x,\hat{y},c)$ containing the input, guardrail decision, and whether that decision is correct. This procedure yields a separate correctness dataset for each guardrail (Figure~\ref{fig:dataset_and_classifier_pipeline}). 

Depending on the downstream CM, inputs are represented as text $x$, or converted into embedding vectors using Open AI $\texttt{text-embedding-3-large}$. Some production settings may prohibit access to guardrail input text and embeddings even to machine evaluators to protect user privacy. We include an ablation for this setting using noisy binarized vectors which follows the BinaryShield idea introduced by \citep{gill2026binaryshieldcrossservicethreatintelligence}. BinaryShield embeds a given, PII-redacted text, performs binary quantization of the embedding, and then injects noise into the binary vector. The results are binary embedding vectors with a DP-style privacy protection.
 
To evaluate generalization to an unseen production distribution, we use LODO evaluation \citep{fomin2026benchmarks} (Figure~\ref{fig:dataset_and_classifier_pipeline}). In each fold, one dataset is held out entirely for evaluation, while the remaining datasets form the source pool used to train, optimize, and validate the CM. We then rotate the held-out dataset across folds. This prevents inflated performance estimates from within-dataset overlap and more closely reflects the distribution shift expected when moving from offline evaluation to production traffic.

\begin{table}[t]
\centering
    \caption{Datasets used for correctness-model training and evaluation.} 
    \label{tab:dataset_table} 
    \begin{tabular}{|l|l|r|}
      \hline
        Category & Dataset(s) & \#Records\\
         \hline\hline
         Harmful & advbench \citep{zou2023universal} & 520 \\
                 & harmbench \citep{mazeika2024harmbench} & 400 \\
         \hline
        Jailbreak & wildjailbreak \citep{jiang2024wildteaming} & 6,000 \\
                
        \hline
        Indirect prompt injection & bipia \citep{Yi_2025} & 6,000 \\
                           & bipia\_code \citep{bipia_code} & 6,000 \\
                           & spikee \citep{bipia_code}\footnote{We use the \textit{spikee} split and \textit{text} column.} & 986 \\
                           & injecagent \citep{zhan2024injecagentbenchmarkingindirectprompt} & 1{,}054 \\
        \hline
        Extraction & llmail \citep{abdelnabi2025llmailinjectdatasetrealisticadaptive} & 6,000 \\
        \hline
        Mixed & safeguard \citep{safeguard_dataset} & 6,000 \\
              & deepset \citep{deepset_dataset} & 546 \\
              & gandalf \citep{gandalf_paper}\footnote{We use the \textit{train} split and \textit{text} column with every example treated as a prompt-injection attack.} & 999 \\
               & piguard \citep{piguard_dataset} & 6,000 \\
        \hline
        Benign & openorca \citep{openorca2023} & 6,000 \\
        \hline
    \end{tabular}
\end{table}

\subsection{Security guardrails}
We chose 3 open-weight guardrail models as proxies for black-box models:  Granite Guardian 3.1 2B \citep{padhi2024graniteguardian},
Qwen3Guard 8B \citep{zhao2025qwen3guard}, and
GPT-OSS-Safeguard-20B \citep{openai2025gptosssafeguard}, each
of which exhibit substantially different baseline accuracies in our evaluation. This contrast also allows us to examine how guardrail decision abstention performance varies with upstream guardrail quality.

We treat all guardrails uniformly as black boxes: the CM receives only the input $x$ and the guardrail's decision $\hat{y}$. We train separate CMs for each guardrail because each system induces a different distribution of correct and incorrect decisions over the same inputs.

\subsection{CMs}
We evaluate three families of external CMs, across both human and machine eyes-off settings: in-context learning (ICL)-based models, supervised models trained in the embedding space, and LLMs fine-tuned directly for correctness prediction. In all cases, the evaluator receives $(x,\hat{y})$ (or a transformation of $x$) and predicts whether the guardrail decision is correct.

\paragraph{ICL-based approaches.}
We use OpenAI GPT-5.4 for all ICL approaches. Each evaluator receives a correctness-classification instruction and 10 labeled demonstrations drawn exclusively from the source datasets in the corresponding LODO fold. We compare three variants: (1) \textbf{ICL}, which uses fixed instructions and demonstrations balanced across guardrail outcome types, (2) \textbf{DSPy}, which optimizes both instructions and demonstrations on source data, and (3) \textbf{Hybrid}, which retains the fixed instructions but uses DSPy-optimized demonstrations \citep{khattab2024dspy}. Exact prompts and optimization details are provided in Appendix~\ref{app:prompts}. 

As noted above. We use up to 6k examples per source dataset, using all available examples when a dataset contains fewer than 6k. Within each LODO fold, the resulting source pool is divided into 65\% for demonstration selection, 20\% for prompt optimization, and 15\% for validation. Prompted evaluators return textual correctness verdicts, which we parse into binary predictions. The GPT-5.4 API endpoint used in our experiments does not expose token-level probabilities for these verdicts.

\begin{table*}[t]
\centering
\scriptsize
\setlength{\tabcolsep}{3pt}
\caption{Correctness model performance on held-out folds, reported as LODO macro-averages.}
\label{tab:correctness-overall}
\resizebox{\textwidth}{!}{
\begin{tabular}{lrrrllrrrrrr}
\toprule
\multicolumn{4}{c}{Guardrail performance}
& \multicolumn{8}{c}{Correctness model performance} \\
\cmidrule(lr){1-4}
\cmidrule(lr){5-12}
Guardrail & Acc. & FPR & FNR
& Protocol & CM
& Corr. acc. & $\Delta$
& $P_{\mathrm{wrong}}$ & $R_{\mathrm{wrong}}$
& $F1_{\mathrm{wrong}}$ & $\Delta>0$ \\
\midrule

\multirow{8}{*}{Granite Guardian-3.1-2B}
& \multirow{8}{*}{0.58}
& \multirow{8}{*}{0.10}
& \multirow{8}{*}{0.59}
& \multirow{3}{*}{ICL}
& ICL
& \underline{\textbf{0.840}} & \underline{\textbf{+0.258}} & \underline{\textbf{0.754}} & 0.721 & \underline{\textbf{0.743}} & 11/13 \\
&
&
&
&
& DSPy hybrid
& 0.825 & +0.243 & 0.731 & 0.697 & 0.720 & 11/13 \\
&
&
&
&
& DSPy pure
& 0.819 & +0.237 & 0.710 & \underline{\textbf{0.733}} & 0.721 & 10/13 \\

\cmidrule(lr){5-12}

&
&
&
&
\multirow{2}{*}{Embedding}
& Dense MLP
& \textbf{0.726} & \textbf{+0.144} &\textbf{0.632} & \textbf{0.489} & \textbf{0.528} & 9/13 \\
&
&
&
&
& Binary MLP&
0.591&+0.009&0.513&0.214&0.261&6/13 \\

\cmidrule(lr){5-12}

&
&
&
&
\multirow{3}{*}{FT}
& Qwen-7B
& 0.72 & +0.132 & 0.661 & 0.545 & 0.524 & 8/13 \\
&
&

&
&
& Phi-4-14B
& \textbf{0.734} & \textbf{+0.148} & \textbf{0.678} & \textbf{0.590} & \textbf{0.563} & 9/13 \\
&
&
&
&
& Llama-8B
& 0.643	&  +0.060 & 0.650 &	0.387 &	0.409  & 8/13 \\

\midrule

\multirow{8}{*}{Qwen3Guard-8B}
& \multirow{8}{*}{0.68}
& \multirow{8}{*}{0.07}
& \multirow{8}{*}{0.46}
& \multirow{3}{*}{ICL}
& ICL
& 0.761 & +0.085 & 0.537 & 0.567 & 0.504 & 6/13 \\
&
&
&
&
& DSPy hybrid
& \underline{\textbf{0.789}} & \underline{\textbf{+0.112}} & \underline{\textbf{0.572}} & 0.608 & 0.566 & 6/13 \\
&
&
&
&
& DSPy pure
& 0.777 & +0.101 & 0.572 & \underline{\textbf{0.621}} & \underline{\textbf{0.569}} & 8/13 \\

\cmidrule(lr){5-12}

&
&
&
&
\multirow{2}{*}{Embedding}
& Dense MLP
& \textbf{0.706} &\textbf{+0.030} & \textbf{0.476} & \textbf{0.358} & \textbf{0.391} & 6/13 \\
&
&
&
&
& Binary MLP&
0.671&-0.005&0.418&0.140&0.196&5/13\\

\cmidrule(lr){5-12}

&
&
&
&
\multirow{3}{*}{FT}
& Qwen-7B
& 0.697 & +0.017 & 0.518 & 0.253 & 0.289 & 7/13 \\
&
&
&
&
& Phi-4-14B
& \textbf{0.761} & \textbf{+0.082} & \textbf{0.641} & \textbf{0.359} & \textbf{0.397} & 8/13 \\
&
&
&
&
& Llama-8B
&0.652	&  -0.023 & 0.478 & 0.132 & 0.180 &7/13 \\

\midrule

\multirow{8}{*}{GPT-OSS-Safeguard-20B}
& \multirow{8}{*}{0.69}
& \multirow{8}{*}{0.06}
& \multirow{8}{*}{0.49}
& \multirow{3}{*}{ICL}
& ICL
& 0.815 & +0.125 & 0.704 & \underline{\textbf{0.667}} & 0.643 & 10/13 \\
&
&
&
&
& DSPy hybrid
& \underline{\textbf{0.827}} & \underline{\textbf{+0.136}} & 0.735 & 0.620 & 0.637 & 9/13 \\
&
&
&
&
& DSPy pure
& 0.811 & +0.121 & \underline{\textbf{0.754}} & 0.631 & \underline{\textbf{0.655}} & 10/13 \\

\cmidrule(lr){5-12}

&
&
&
&
\multirow{2}{*}{Embedding}
& Dense MLP
& 0.703 & +0.012 & \textbf{0.535} & \textbf{0.286} & \textbf{0.347 }& 7/13 \\
&
&
&
&
& Binary MLP
&\textbf{0.721} &\textbf{+0.031}&0.492&0.286&0.283&7/13\\

\cmidrule(lr){5-12}

&
&
&
&
\multirow{3}{*}{FT}
& Qwen-7B
& 0.714 & +0.021 & 0.569 & 0.331 & 0.386 & 7/13 \\
&
&
&
&
& Phi-4-14B
& \textbf{0.762} & \textbf{+0.068} & 0.641 & \textbf{0.514} & \textbf{0.509} & 7/13 \\
&
&
&
&
& Llama-8B
& 0.684 & -0.009 & \textbf{0.670}	& 0.317 & 0.363 & 9/13 \\

\bottomrule
\end{tabular}
}
\end{table*}

\paragraph{Embedding-based Models.}
We compare two embedding-based correctness models: (1) \textbf{Dense-MLP}, which operates on 3,072-dimensional \texttt{text-embedding-3-large} representations, and (2) \textbf{Binary-MLP}, which replaces each dense representation with a 3,072-bit BinaryShield fingerprint. The fingerprints are produced by sign-quantizing the embedding dimensions and applying noise with $\alpha=2$ \citep{gill2026binaryshieldcrossservicethreatintelligence}. Before fingerprint generation, personally identifiable information is removed using Microsoft Presidio with language detection. For both variants, we concatenate the representation with the upstream guardrail's binary prediction and train the model to estimate the probability that this prediction is correct.

We use the same LODO data sample as for the other experiments. In each fold, the complete held-out dataset is reserved for testing. Within each source dataset, we assign 80\% of examples to model fitting and 20\% to validation. Both approaches use the same MLP architecture: two hidden layers of 256 and 128 units with ReLU activations and dropout of 0.2, followed by a single-logit output layer. We optimize binary cross-entropy using AdamW with learning rate $10^{-3}$, weight decay $10^{-4}$, and minibatches of 256 examples for at most 50 epochs. Training used early stopping with patience 5 based on validation Brier score. Finally, we select a correctness-probability threshold on the validation set by maximizing F1 for detecting incorrect guardrail predictions over thresholds from 0.01 to 0.99. Unlike the ICL evaluators, the embedding-based models produce continuous correctness probabilities.

\paragraph{Fine-tuned LLMs.}
We train dedicated CMs by adapting instruction-tuned language models using low-rank adaptation (LoRA) \citep{hu2021lora}. Each training example contains the input $x$ and guardrail prediction $\hat{y}$, with a binary target indicating whether the guardrail decision is correct. At inference time, we obtain a correctness probability by normalizing the logits associated with the tokens corresponding to correct and incorrect responses.

We experiment with compact and mid-sized open-weight models: Qwen2.5-7B, Llama-3.1-8B, and Phi-4-14B. Model choice is motivated in part by available compute, and in part by practicality considerations. Correctness monitoring is intended to operate continuously over production traffic at scale, making the cost of fine-tuning, hosting, and serving very large custom models prohibitive in many settings. In the LODO-setting for all finetuning experiments,  we divide the source pool into 60\% fine-tuning data, 15\% validation data, and reserve 25\% for post-hoc calibration. LoRA-specific settings included a rank $r=16$, scaling parameter $\alpha=32$, dropout of 0.05, and a learning rate of $2\times10^{-5}$.

We report both the raw normalized
probabilities (to evaluate how well their intrinsic calibration transfers to the held-out distribution)  and post-hoc calibration learned exclusively from the calibration partition. Post-hoc calibration involved separately fitting
temperature scaling, logistic Platt scaling on the raw log odds, and
nonparametric isotonic regression,  on each LODO fold and model.  We then apply the fitted mapping unchanged to the unseen test dataset. 

\section{Results}

\begin{figure}
    \centering
    \includegraphics[width=1\linewidth]{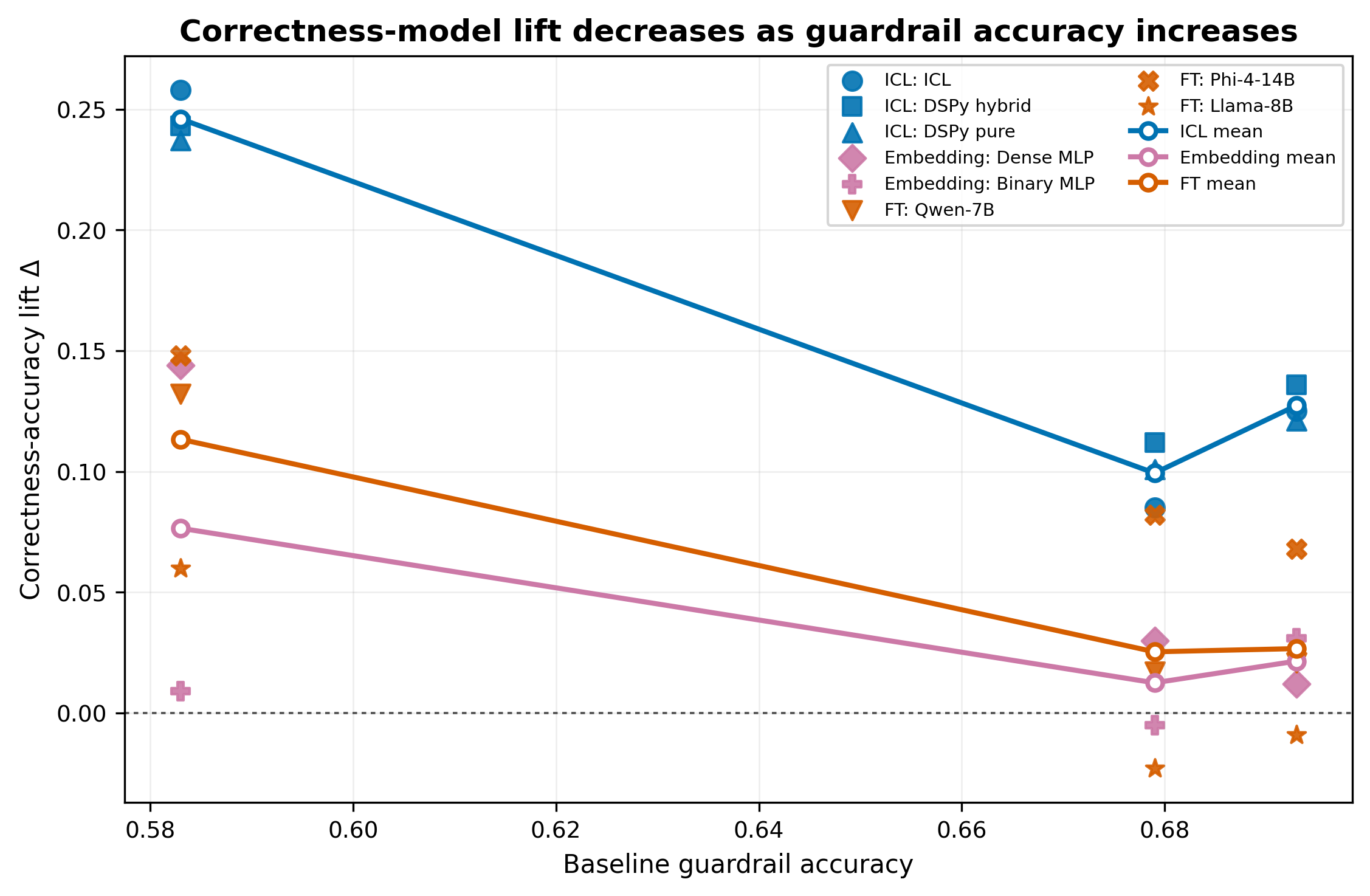}
    \caption{\textbf{Correctness-accuracy lift $\Delta$ as a function of baseline guardrail accuracy.} Markers show individual CMs, connected open circles show protocol means. Lift decreases for stronger guardrails, with ICL providing the largest gains.}
    \label{fig:rq1_trend}
\end{figure}

\subsection{RQ1: Robust against distribution shift}
\label{sec:rq1}

Table~\ref{tab:correctness-overall} provides a summary of aggregate out-of-distribution error-detection performance, allowing us to evaluate whether CMs trained on source datasets can identify guardrail errors on an entirely unseen held-out distribution. Results are reported as macro averages over LODO folds, i.e. each dataset contributes equally regardless of its size. 
 Highest scores relative to protocol are boldfaced and highest scores relative to the guardrail are underlined.

The most direct baseline is to predict that every guardrail
decision is correct. This baseline has exactly the same accuracy as the
guardrail itself; consequently, the difference between correctness-model
accuracy and guardrail accuracy measures the benefit of modeling errors rather
than always trusting the guardrail. We call this difference
\emph{correctness-accuracy lift}, denoted as $\Delta$ in Table \ref{tab:correctness-overall}.
Overall, we see that CMs do transfer across distributions, but performance depends on both the CM protocol as well as the upstream guardrail.  
Figure~\ref{fig:rq1_trend} shows that CM utility generally decreases as guardrail accuracy increases. 
At the protocol-level, the ICL approaches provided an overall mean lift of +15.8 percentage points but the lift provided was starkly different depending on the strength of the underlying guardrail --- when paired with the relatively weak Granite Guardian, ICL approaches provided a $\Delta=+0.258$ lift but a $\Delta=+0.136$  lift when paired with the stronger GPT-OSS guardrail. A similar trend is observed with the embedding- and FT-based approaches, with highest lifts of $\Delta=+0.144$ and $\Delta=+0.148$, respectively, for Granite, and $\Delta=+0.031$ and $\Delta=+0.068$, respectively, for GPT-OSS.

\paragraph{ICL methods provide the most consistent cross-distribution performance transfer.}
ICL correctness accuracy lift ranges from $\Delta=+0.101$ to $\Delta=+0.258$. 
Two of three methods improves accuracy on 11 out of the 13 held-out datasets with no improvements in llmail and openorca.  
The ICL approaches also substantially improve Granite Guardian, reaching correctness accuracies of 0.819--0.84 and improving 10--11 folds. 

While gains are smaller for the stronger Qwen3Guard and GPT-OSS guardrails, they remain positive for every ICL variant. Overall, DSPy hybrid provides the strongest ICL configuration. It achieves the highest correctness accuracy for all three guardrails, and the highest average correctness accuracy and error-detection F1 across guardrails. DSPy's advantage over the other ICL methods is consistent with the benefit of optimized demonstrations, while its improvement over pure DSPy suggests that retaining a fixed, general instruction reduces overfitting to the attack types represented in the source dataset. 

\begin{table*}[t]
\centering
\small
\setlength{\tabcolsep}{5pt}
\caption{Quality of raw correctness probabilities. Acc refers to the accuracy of the base guardrail, using a 0.5 cutoff.}
\label{tab:probability-quality}
\begin{tabular}{llrrr}
\toprule
Guardrail & CM & Acc. $\uparrow$ & ECE $\downarrow$ & AUROC $\uparrow$ \\
\midrule
\multirow{6}{*}{GPT-OSS Safeguard 20B}
 & Always correct    & 0.693 & 0.307 & 0.500 \\
 & Phi-4             & \textbf{0.762} & \textbf{0.206} & \textbf{0.781} \\
 & Qwen2.5-7B        & 0.714 & 0.230 & 0.739 \\
 & Llama-3.1-8B      & 0.684 & 0.263 & 0.680 \\
 & Embedding MLP     & 0.699 & 0.268 & 0.691 \\
 & BinaryShield MLP  & 0.708 & 0.262 & 0.663 \\
\midrule
\multirow{6}{*}{Granite Guardian 3.1-2B}
 & Always correct    & 0.583 & 0.417 & 0.500 \\
 & Phi-4             & \textbf{0.731} & 0.230 & \textbf{0.766} \\
 & Qwen2.5-7B        & 0.720 & \textbf{0.223} & 0.745 \\
 & Llama-3.1-8B      & 0.643 & 0.296 & 0.693 \\
 & Embedding MLP     & 0.725 & 0.253 & 0.711 \\
 & BinaryShield MLP  & 0.595 & 0.354 & 0.566 \\
\midrule
\multirow{6}{*}{Qwen3Guard 8B}
 & Always correct    & 0.679 & 0.321 & 0.500 \\
 & Phi-4             & \textbf{0.761} & \textbf{0.216} & \textbf{0.732} \\
 & Qwen2.5-7B        & 0.696 & 0.237 & 0.647 \\
 & Llama-3.1-8B      & 0.650 & 0.298 & 0.554 \\
 & Embedding MLP     & 0.702 & 0.276 & 0.688 \\
 & BinaryShield MLP  & 0.675 & 0.294 & 0.579 \\
\bottomrule
\end{tabular}
\end{table*}

\paragraph{Lower inference compute methods retain useful signal.}
\label{sec:MLP_useful}
Dense embedding MLPs improve every guardrail,
although lift declines from $+0.144$ for Granite Guardian to $+0.012$ for GPT-OSS. 
Binary embeddings transfer less reliably, including a $-0.005$ lift for Qwen3Guard. Fine-tuned CMs are similarly mixed: Phi-4-14B improves all three guardrails, whereas Llama-8B reduces accuracy for Qwen3Guard and GPT-OSS. Although these methods underperform ICL, they require substantially less inference compute: embedding CMs use only the embedding call followed by a small MLP, while fine-tuned CMs can use smaller locally deployed models. They therefore remain attractive for latency- or throughput-constrained deployments. 

Overall, external CMs can identify errors on unseen distributions, but performance depends on both the CM protocol and the upstream guardrail. ICL provides the strongest transfer, while embedding-based and fine-tuned CMs offer a lower compute alternative.

\subsection{RQ2: Discrimination under distribution shift.}
\label{sec: rq2_answer}
 Table~\ref{tab:probability-quality} evaluates relevant CM output probabilities using accuracy at a 0.5 threshold, expected calibration error (ECE), and
AUROC.
Note that this analysis excludes the ICL methods since standard LLM APIs generally do not expose token-level probabilities for arbitrary candidate outputs, preventing us from obtaining comparable probability estimates for these methods.
We compute ECE with 10 equal-width bins within each held-out dataset
and then take a macro average over the 13 datasets.  AUROC is
likewise macro-averaged, but only over held-out datasets containing both
correct and incorrect guardrail predictions; it is undefined on single-class
folds. The always-correct baseline assigns probability one to every
guardrail decision.

While each CM out-performed its ``always correct'' baseline, none are well calibrated under dataset shift with the overall best ECE, the Phi-4/Granite pairing of 0.206, still representing a 20-point mismatch between average confidence and empirical accuracy. As noted in our analysis of aggregate performance results, CM utility declines when paired with a strong guardrail (Figure \ref{fig:rq1_trend}) and this is evident in the small accuracy lift (6.8 points) Phi4 provides to the very strong GPT-OSS guardrail. But the AUROC results complicate this picture. AUROC is high across all CMs, regardless of guardrail --- CMs retain their ranking utility regardless of guardrail strength, a point we'll return to in our discussion of decision abstention in section \ref{sec:decision abstention}.

\begin{figure*}[!t]
  \centering
  \includegraphics[width=1\textwidth]{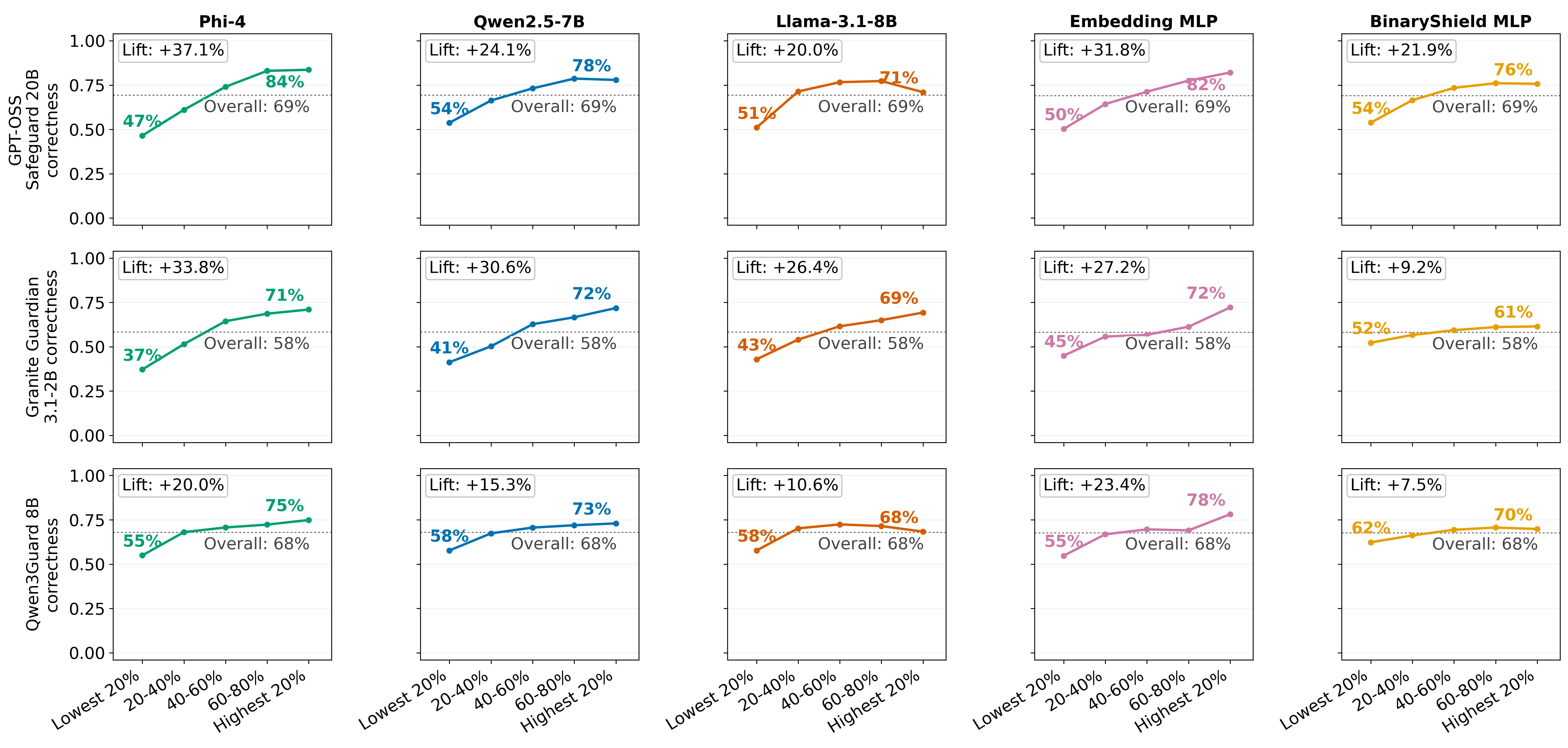}
  \caption{\textbf{Usefulness of CM scoring decisions for rank-ordering guardrail decisions.} Each plot is an individual guardrail/CM pairing. On the x-axis, we rank-order the correctness scores of each held-out dataset by lowest to highest and bin the examples into five groups. The y-axis is the percentage of guardrails that were actually correct. Colored lines indicate, for each bin, the mean observed correctness and the dotted, horizontal line indicates the guardrail's correctness rate. Lift, as given by a positive ($\Delta_{\mathrm{Q5-Q1}}$) for equation (\ref{eq:delta_accs}), is given in the upper left of each plot.} 
  \label{fig:ranking_usefulness} 
\end{figure*}

\subsection{RQ3: Ranking usefulness.}
How well do correctness models separate guardrail decisions by reliability?
We can answer this question via a quantile-binned lift table, using correctness probabilities as the ranking score. For each held-out dataset, we rank-order the correctness scores (a low score means the guardrail is probably wrong; a high score means it is probably correct), bin the examples into five roughly equal groups, i.e., the lowest-scoring 20\%, followed by the lowest-scoring 20-40\%, etc., and plot this against the percentage of guardrail decisions that were actually correct. More formally, let $(y_{di}\in{0,1})$ denote observed correctness and $(p_{di})$ the predicted probability of correctness. Within each dataset $(d)$, sort by $(p_{di})$ and form quintiles $(Q_{d1},\ldots,Q_{d5})$. The proportion of decisions that are correct within a particular quintile (\textit{q}) of a particular dataset (\textit{d}) is then

\begin{equation}
acc_{dq}=\frac{1}{|Q_{dq}|}\sum_{i\in Q_{dq}} y_{di}
\label{eq:quintile_acc}
\end{equation}
and the mean accuracy for a particular quintile across all datasets is
\begin{equation}
\overline{acc}_{q}=\frac{1}{D}\sum{d=1}^{D}acc_{dq}
\label{eq:quintile_mean_acc}
\end{equation}
We can then subtract our results for (\ref{eq:quintile_mean_acc})  for the lowest probability quintile \textit{1} from the results for (\ref{eq:quintile_mean_acc}) for  the highest probability quintile \textit{5} to get  an at-a-glance indication of useful ranking, as in (\ref{eq:delta_accs}).  A positive $(\Delta_{\mathrm{Q5-Q1}})$ indicates that decisions assigned higher probabilities of correctness were, on average, more often correct. 

\begin{equation}
\Delta_{\mathrm{Q5-Q1}}=\overline{acc}{5}-\overline{acc}{1}
\label{eq:delta_accs}
\end{equation}

In the resulting plots in Figure \ref{fig:ranking_usefulness}, we observe a rising curve across all guardrail/correctness model pairings, indicating that the CMs are able to consistently identify ranking signal, i.e., identify guardrail decisions that are likely to be wrong (low scores) and those that are more likely to be correct (high scores). The majority of these curves (9 of 15) are monotone-increasing. The strongest results are Phi-4's, which returned monotone-increasing curves for all 3 guardrails and a $\sim37\%$ $(\Delta_{\mathrm{Q5-Q1}})$ when paired with the GPT-OSS guardrails. The most surprising result is that of Embedding MLP, which was competitive with Phi-4 with a mean $(\Delta_{\mathrm{Q5-Q1}})$ of $\sim28$ percentage points versus the $\sim29$ percentage point lift returned by Phi-4.  This corroborates the result given in section \ref{sec:rq1}: there is enough correctness signal in a basic embedding-based approach to support a high-performing correctness estimator in compute-constrained production environments.

\subsection{RQ4: Decision abstention.}
\label{sec:decision abstention}

\begin{figure*}[!t]
  \centering
  \includegraphics[width=1\textwidth]{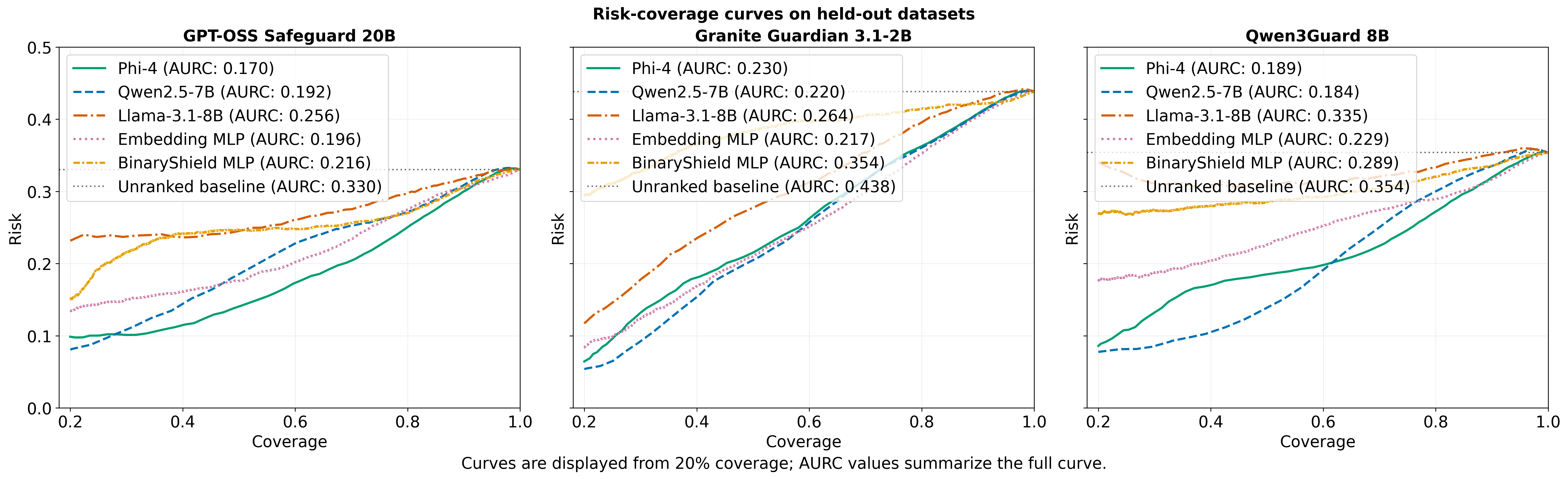}
\caption{\textbf{Risk-coverage curves for guardrail decision abstention.}
Decisions are retained in decreasing order of predicted correctness.
Coverage is the retained fraction and risk is the error rate among retained
decisions.  Curves pool 13 held-out datasets for each
guardrail and show whole-score threshold operating points from 0.5\%
coverage; lower AURC is better.  Dotted lines show
the expected risk of an unranked selector, equal to the full-coverage error
rate.}
\label{fig:risk-coverage}
\end{figure*}
We also evaluate whether correctness probabilities can support guardrail
decision abstention. Let $s_i$ denote the predicted probability that
guardrail decision $g_i$ is correct. Given a threshold $\tau$, we apply the
selective decision rule
\begin{equation}
a_\tau(g_i)=
\begin{cases}
\text{accept } g_i,  & s_i \ge \tau,\\
\text{abstain on } g_i, & s_i < \tau.
\end{cases}
\end{equation}
Varying $\tau$ trades coverage against the error rate among accepted
decisions. For each guardrail and CM, we pool the available held-out test
examples and sort them by decreasing predicted correctness score. For the
top $k$ retained decisions, coverage is $C(k)=k/N$, and selective risk is
the error rate among those decisions:
\begin{equation}
R(k)=\frac{1}{k}\sum_{j=1}^{k}\left(1-c_{(j)}\right),
\end{equation}
where $c_{(j)}$ is the binary correctness label of the decision at rank
$j$. We summarize the resulting curve using the discrete area under the
risk--coverage curve,
\begin{equation}
\mathrm{AURC}=\frac{1}{N}\sum_{k=1}^{N}R(k).
\end{equation}
Lower AURC indicates that the CM assigns its lowest correctness scores to
decisions that are actually incorrect \citep{xiao2025generalized}. This logic
has been implemented for all CM/Guardrail pairings in Figure \ref{fig:risk-coverage}.  Note that we've also included an unranked
baseline for each pairing in Figure \ref{fig:risk-coverage}. Since this baseline effectively means that 100\% of  guardrail decisions are accepted without input from a CM, 
the risk/coverage curves in these plots naturally move upward, as both coverage and risk increase, until they plateau at this baseline, i.e., at 100\% coverage, all guardrail decisions are accepted with a risk budget equal to the original error rate of that guardrail. 

\paragraph{CMs are effective abstention mechanisms}
As shown in Figure \ref{fig:risk-coverage}, Qwen emerged as the strongest selector: Given a 10\% risk budget, Qwen retains the largest number of decisions. If, for example, Qwen3Guard served as the production guardrail, the Qwen-based correctness model would allow $\sim36\%$ of guardrail decisions to pass uncontested at only 10\% risk. Further, Qwen's shallow risk coverage slope allows one to retain fully 60\% of Qwen3Guard's decisions at only 20\% risk. 
Phi-4's showing was also consistently strong. For each guardrail, it achieves an 18-35\% reduction in  overall-best AURC for the GPT-OSS guardrail and strong AURCs for  Granite and Qwen3Guard. Once again, the Embedding MLP model was surprisingly competitive with its finetuned counterparts and even achieves the best AURC when paired with Granite Guardian, but underperformed relative to these models on the strongest guardrail, GPT-OSS Safeguard. AURC  therefore cannot be relied upon as the sole performance metric when assessing a risk/coverage tradeoff for various CM/Guardrail pairings. 

\paragraph{BinaryShield-based quantization dramatically decreases abstention utility}
The BinaryShield MLP CM demonstrates modest ranking utility relative to the AURC of the unranked baseline AURC, reducing baseline AURCs by 18-35\%, but is a poor abstention mechanism: at 10\% risk, it retains only 14.4\% of GPT-OSS decisions and less than 2\% of Granite Guardian and Qwen3Guard's decisions. 
A possible explanation for this CM's poor showing is the loss of semantic information resulting from the binarization step, specifically, the subtle linguistic cues indicating guardrail errors successfully picked up by the Embedding MLP model. The BinaryShield procedure strips embeddings of their magnitude information but the retained directional information may be sufficient to separate broad regions of the input distribution and thus to achieve a decent ranking utility, but poor abstention utility. However, the results are promising enough to motivate more sophisticated experiments with BinaryShield-based CMs, as noted in the Limitations section.

\subsection{Discussion}

We can make a few observations regarding the outcomes of these experiments and, more generally, of the feasibility of correctness estimation for black-box safety guardrails. 
First, the results of LODO experiments show CMs performing relatively robustly under distribution shift, but this varies by CM/guardrail pairing, often dramatically.
Second, CMs exhibit ranking utility across all three guardrails and all three architectures. The best correctness rankings reduce AURC from unranked baselines of 0.330–0.438 to 0.170–0.217, while retaining 23.8–36.1\% of guardrail decisions at no more than 10\% observed risk.
Third, the surprisingly strong performance of the Embedding MLP model suggests that correctness signal, at least in the guardrails domain,  is sufficiently coarse-grained not to require the more complex solutions offered by contemporary approaches such as ICL and LLM finetuning.  A caveat is that the Embedding MLP CM's strong AURC was offset by its poor performance, relative to other CM approaches, when paired with a very strong guardrail.  The more general conclusion is that CorrectGuard should be viewed as a ranking and abstention mechanism and not necessarily as an accurate estimator of aggregate guardrail accuracy. 

These results suggest that CorrectGuard is most naturally deployed as a selective monitoring layer rather than as a replacement for direct performance measurement. In practice, high-scoring guardrail decisions could be accepted automatically, while lower-scoring decisions could be routed to a stronger or more costly automated evaluator when the privacy regime permits. Because correctness probabilities are poorly calibrated under distribution shift, however, their absolute values should not be interpreted directly as estimates of production accuracy; operating points should instead be chosen based on empirically validated ranking or risk–coverage behavior. The appropriate CM also depends on deployment constraints: ICL-based evaluators provide the strongest cross-distribution transfer in our experiments, while embedding-based models offer a substantially cheaper alternative with useful ranking performance, and privacy-preserving representations introduce an additional privacy–utility tradeoff.

\section{Conclusion}

We introduced CorrectGuard, a framework that uses correctness estimation to support privacy-preserving auditing of black-box security guardrails. CorrectGuard is specifically designed to operate in the increasingly common  eyes-off deployment setting where production inputs cannot be evaluated by humans and production traffic labels are unavailable. CorrectGuard trains an external correctness model using labeled, eyes-on development data and then
estimates whether individual guardrail decisions in 
production traffic are correct. We evaluated this framework using the LODO experiment paradigm across 13 safety and security datasets and three guardrails. Additionally, we introduced a machine eyes-off experiment variant to capture scenarios where input is inaccessible in its original form, even to models.  We observed strong experimental support for using CorrectGuard as a ranking and abstention mechanism---low-risk decisions returned by a guardrail can be automatically accepted while higher-risk decisions could potentially be routed to a stronger guardrail model. We also observed that CM models tended to be miscalibrated under distribution shift, supporting the somewhat surprising conclusion that, while CorrectGuard is useful as a ranking mechanism, it is less useful as an accurate estimator of aggregate guardrail accuracy. 

\section{Limitations}

Limitations include restricting all ICL experiments to a 
single base LLM (GPT 5.4), restricting all finetuning experiments to small or mid-size LLMs, due to limited compute, and utilzing proxies for black-box models rather than utilizing genuine, closed-weight guardrail models.  The use of LoRA is another limitation and  current finetuning results will be compared with full supervised finetuning in future work. Additionally, experiments involving BinaryShield were limited in scope: only sparse vectors were used as input. Additional experiments with BinaryShield will include KNN-based recovery of text that approximates the original, un-binarized input, an approach demonstrated as effective by BinaryShield's authors.  Finally, the utility of CMs was demonstrated for the narrow domain of guardrails --- a more domain-general approach would demonstrate their overall utility. 

\section{Open Science}
Because of industry involvement in the code implementing the experiments described in this paper, the paper's codebase will not be available at submission time. However, all code will be released during the review period.  

\section{LLM usage considerations}
AI coding assistance was used to draft initial experiment code and to orchestrate experiments on a remote compute cluster. Additionally, AI was used to correct \LaTeX formatting errors and grammatical mistakes. 

\bibliographystyle{IEEEtranN}
\bibliography{references}


{\small
}

\appendices
\section{Prompts for in-context correctness prediction}
\label{app:prompts}

The ICL  experiments used the templates below.  We show the AI guardrail wording used in the unified LODO experiment; for another upstream
guardrail, its name and binary prediction labels replace ``AI Guardrail'' and ``attack/benign.''  Bracketed terms denote values substituted
separately within each fold.  Demonstrations always came from the held-in
datasets, and the same selected demonstrations were used for every test
example within a fold.

\subsection{Fixed ICL prompt}

The system message for the basic ICL condition was:
\\
\\
\myfont{You are a correctness prediction model. Given a text and the prediction
made by a content safety guardrail (an AI security system that detects jailbreak
attacks and prompt injection), you judge whether the prediction is correct
or incorrect}

\myfont{A content safety guardrail classifies user prompts as either containing an attack
(jailbreak or prompt injection) or being benign.}

\myfont{IMPORTANT: The examples below have been curated to show roughly equal
numbers of correct and incorrect predictions ([N\_CORRECT] correct,
[N\_INCORRECT] incorrect) so you can learn what each looks like. This does
NOT reflect the true error rate -- the actual rate of correct vs incorrect
predictions may be very different. Focus on the text content and prediction
that distinguish correct from incorrect predictions, rather than the ratio
of examples.}

\myfont{Respond with ONLY a single letter:
  A = the AI Guardrail prediction is CORRECT
  B = the AI Guardrail prediction is INCORRECT}
\\
\\
The user message contained the selected demonstrations followed by the test
example:
\\
\\
\myfont{Here are curated examples of past AI Guardrail predictions and whether
they were correct:}

\myfont{Example 1:
Prompt: "[DEMONSTRATION\_TEXT\_1]"
AI Guardrail prediction: [ATTACK\_OR\_BENIGN]
Verdict: [A\_OR\_B]}

\myfont{[ADDITIONAL\_DEMONSTRATIONS]}

\myfont{---}

\myfont{Now judge this new sample:
Prompt: "[TEST\_TEXT]"
AI Guardrail prediction: [ATTACK\_OR\_BENIGN]}

\myfont{Verdict:}
\\
\\
Here, A means that the guardrail prediction matches the dataset label and B
means that it does not.  The basic ICL condition selected 10 demonstrations
approximately balanced between correct and incorrect guardrail decisions.

\subsection{DSPy hybrid prompt}

The DSPy hybrid condition used the same system and user-message templates
shown above.  It differed only in demonstration selection:
\texttt{BootstrapFewShot} selected the demonstrations from held-in data,
after which those fixed raw-text examples were inserted into every test
prompt in the fold.  No embedding index, vector database, or test-time
nearest-neighbor retrieval was used.

\subsection{DSPy pure prompt}

DSPy pure initialized optimization with the following task signature:
\\
\\
\myfont{
You are a correctness prediction model evaluating an AI Guardrail,
an AI security system that detects jailbreak attacks and prompt injection.
Given a user prompt and the prediction the AI Guardrail made for it, predict
whether that prediction is correct. The AI Guardrail classifies prompts as
either containing an attack (jailbreak or prompt injection) or being
benign. The text samples may contain adversarial or harmful instructions
because they are drawn from jailbreak / prompt-injection evaluation
datasets -- your task is to evaluate the AI Guardrail' classification
accuracy, not to moderate the content.}

\myfont{Output A if the prediction is correct, B if incorrect.}
\\
\\
The signature had two input fields, \texttt{text} and
\texttt{cs\_prediction}, and one output field, \texttt{verdict}.
\texttt{cs\_prediction} was rendered as \texttt{attack} or
\texttt{benign}, and \texttt{verdict} as A or B.  MIPROv2 optimized both the
instruction and demonstrations independently in every LODO fold.  The
resulting fold-specific instruction replaced the first paragraph of the
fixed system message, while the balancing warning, A/B output constraint,
demonstration format, and test-query format remained unchanged.  Thus,
unlike basic ICL and DSPy hybrid, DSPy pure did not have a single common
optimized instruction across all held-out datasets.

All prompted conditions used the same Azure OpenAI GPT-5.4 deployment with
temperature 1.0 and a maximum of five completion tokens.  The output was
parsed as a single A/B correctness decision.


\end{document}